\documentclass[11pt]{article} 

\usepackage{amsmath,amssymb,amsbsy,amsfonts,latexsym,graphicx}
\usepackage{hyperref}
\usepackage{cite}
\usepackage{setspace}
\usepackage{tikz}
\usepackage{caption}
\usepackage{xcolor}

\renewcommand{\a}{\alpha}      \renewcommand{\b}{\beta}
\renewcommand{\d}{\delta}

\renewcommand{\th}{\theta}

\newcommand{\m}{\mu}         
\newcommand{\n}{\nu}

\newcommand{\cala}{\mbox{${\cal A}$}}

 \newcommand{\calh}{\mbox{${\cal H}$}}

\newcommand{\calo}{\mbox{${\cal O}$}}

\newcommand{\nn}{\nonumber}

\newcommand{\tr}{{\rm tr}\,}

\newcommand{\wt}{\widetilde}

\allowdisplaybreaks

\renewcommand{\thefootnote}{\arabic{footnote}}
\newcommand{\be}{\begin{equation}}
\newcommand{\ee}{\end{equation}}

\begin{document}

\begin{titlepage}
	\bigskip

	\bigskip\bigskip

	\bigskip
	
	\begin{center}
		
		\centerline
		{\LARGE \bf {Global symmetry violation from non-isometric codes}}
		\bigskip
		
		\bigskip
		{\Large \bf { }} 
		\bigskip
		\bigskip
	\end{center}

	\begin{center}

			\renewcommand{\thefootnote}{\fnsymbol{footnote}}
		
		{\large
			Jong-Hyun Baek$^a$\,\footnote[1]{\texttt{\url{jonghbaek@gmail.com}}},
			Kang-Sin Choi$^{a, b}$\,\footnote[2]{\texttt{\url{kangsin@ewha.ac.kr}}},
			
		}
		
		\renewcommand{\thefootnote}{\arabic{footnote}}
		
		\bigskip \rm
		
		\bigskip
		$^a$ Institute of Mathematical Sciences, Ewha Womans University, Seoul 03760, Korea 
		
		\rm 
		\bigskip
		$^b$ Scranton Honors Program, Ewha Womans University, Seoul 03760, Korea\\
		\rm
		\bigskip

		\bigskip \rm
		\bigskip
		
		\rm

		\bigskip
		\bigskip

	\end{center}

	\bigskip\bigskip
	\begin{abstract}
We study the no-global-symmetry conjecture in quantum gravity by modeling black holes as non-isometric codes that encode the interior states with global charges into the fundamental states. The fluctuation in the inner products of the charged states can be larger compared to the case without charges. The non-isometric map causes states with different charges to have non-zero overlaps, signaling symmetry violation. The Renyi entropies of the radiation with global charges are found to be consistent with the quantum extremal surface formula. We compute various forms of Renyi relative entropy, as well as the fidelity, to quantify the degree of a global symmetry violation in Hawking radiation, demonstrating that global symmetries are indeed violated. We also comment on the instability of the black hole remnant.

		\medskip
		\noindent
	\end{abstract}
	\bigskip \bigskip \bigskip

	\vspace{1cm}

	\vspace{2cm}
	
\end{titlepage}

\tableofcontents


\section{Introduction}

Global symmetries are known to be inconsistent with quantum gravity. The absence of global symmetries has long been conjectured based on various arguments \cite{Giddings:1988cx, Kallosh:1995hi, Banks:2010zn}, including holographic ones in the context of AdS/CFT \cite{ Harlow:2018jwu, Harlow:2018tng, Harlow:2020bee, Heckman:2024oot}. Advances in understanding black hole physics and replica wormholes \cite{Penington:2019npb, Almheiri:2019psf, Almheiri:2019hni,Penington:2019kki, Almheiri:2019qdq, Almheiri:2020cfm} have also provided quantitative insights into how global symmetries might be violated in quantum gravity \cite{Chen:2020ojn, Hsin:2020mfa, Bah:2022uyz}. Specifically, it was pointed out in \cite{Chen:2020ojn} that the relative entropy between the exact density matrix of the radiation and the density matrix transformed by a global symmetry operator can signal the violation of conservation of global charge, capturing the effects of charged particles flowing out through replica wormholes. 

An important development in the emergence of spacetime came from the study of quantum error correction, where the bulk degrees of freedom in the effective field theory are mapped into the boundary degrees of freedom by an isometric map, preserving the inner product \cite{Almheiri:2014lwa, Dong:2016eik, Harlow:2016vwg}. A further proposal has been made that non-isometric codes can describe black hole interior, where the number of effective degrees of freedom can be much larger than the horizon area \cite{Akers:2022qdl}. It implies that a large set of states in the effective description of black hole interior must be annihilated by the holographic map to the fundamental description, which incorporates quantum gravity effects. The non-isometric nature of the map was crucial in obtaining results that are consistent with the quantum extremal surface (QES) formula. In this paper, we use non-isometric maps to show that global symmetries are violated in quantum gravity by computing the Renyi entropy and relative entropy. 

In section \ref{two}, we introduce the non-isometric codes when global charges are included. The fluctuation of the inner product between states in the fundamental and effective descriptions is shown to be bounded. In section \ref{three}, the Renyi entropies are computed to derive the QES formula and the Renyi relative entropy is calculated to demonstrate global symmetry violation. We also discuss how remnant-like states could contribute to the entropy but their contribution is eventually suppressed at late times. Appendix \ref{appA} contains the computation of the upper bound for the fluctuation of inner product. Appendix \ref{singleu} discusses an alternative model of single unitary map.

\section{Non-isometric codes}\label{two}

Following the work of Akers et al. \cite{Akers:2022qdl}, we model a black hole as a non-isometric map that encodes the effective degrees of freedom in the black hole interior to the fundamental degrees of freedom.  
In the fundamental description, the black hole is described as a quantum system $B$.
We do not need to know its details, except the following. (1)  It evolves unitarily and (2) the dimension $|B|$ of the Hilbert space $\calh_B$ of $B$ decreases as the black hole evaporates. The entropy of the black hole is $S_{N}=\log |B|$.

In the effective description, gravity is classical and matters are treated by quantum field theory. The semiclassical entropy for the Hawking radiation exceeds the area of the black hole horizon at late times. This implies that the effective degrees of freedom in the black hole interior are much larger than its size. So the number of the fundamental degrees of freedom in the black hole interior must be reduced. This leads to the idea that black holes can be thought of as non-isometric codes. 

The black hole interior degrees of freedom consist of the left-movers $\ell$ and the right-movers $r$ with their Hilbert spaces $\calh_\ell$ and $\calh_r$, respectively. States in $\calh_\ell$ account for the degrees of freedom from which the black hole was made. The right-movers $r$ are entangled with the radiation $R$ in a Hawking radiation state. An observer outside of the black hole has access only to the radiation $R$.

We introduce a linear map $V$ from the effective description to the fundamental description  
\begin{equation}
	V:\calh_\ell\otimes\calh_r \rightarrow \calh_B 
	.
\end{equation}
and add some extra system $f$ and $P$ such that
\begin{equation}
	\calh_\ell\otimes\calh_f\otimes\calh_r=\calh_B\otimes\calh_P,  \qquad |\ell||f||r|=|B||P|.
\end{equation}
The effective degrees of freedom are expected to be much larger than the fundamental degrees of freedom at late times. So we take $|P|\gg 1$.

\begin{figure}[t]\begin{center}
		\begin{tikzpicture}
			\draw (0,0) node {$V = \sqrt{|P|}$};
			\draw (1,0.7)--(1,-0.7)--(3,-0.7)--(3,0.7)--(1,0.7);
			\draw (2,0) node {$U$};
			\draw (1.3,-0.7)--(1.3,-1.7);
			\draw (2,-0.7)--(2,-1);		
			\draw (2.7,-0.7)--(2.7,-1.7);
			\draw (1.3,-1.7) node[below] {$\ell$};
			\draw (2,-1) node[below, inner sep=0mm] {\small $|\psi_0\rangle_f$};
			\draw (2.7,-1.7) node[below] {$r$};
			\draw (1.5,0.7)--(1.5,1.5);
			\draw (2.5,0.7)--(2.5,1.5);
			\draw (1.5,1.5) node[above] {$B$};
			\draw (2.5,1.5) node[above, inner sep=0.5mm] {$\langle 0|_P$};
		\end{tikzpicture}
		\qquad\qquad
		\begin{tikzpicture}
			\draw (0,0) node {$W = \sqrt{|P_G|}$\,\,\,};
			\draw (1,0.7)--(1,-0.7)--(3,-0.7)--(3,0.7)--(1,0.7);
			\draw (2,0) node {$U_G$};
			\draw (1.3,-0.7)--(1.3,-1.7);
			\draw (2,-0.7)--(2,-1);		
			\draw (2.7,-0.7)--(2.7,-1.7);
			\draw (1.3,-1.7) node[below] {$Q_\ell$};
			\draw (2,-1) node[below, inner sep=0mm] {\small $|\wt{\psi}_0\rangle_{f_G}$};
			\draw (2.7,-1.7) node[below] {$Q_r$};
			\draw (1.5,0.7)--(1.5,1.5);
			\draw (2.5,0.7)--(2.5,1.5);
			\draw (1.5,1.5) node[above] {${C}$};
			\draw (2.5,1.5) node[above, inner sep=0.7mm] {$\langle 0|_{P_G}$};
		\end{tikzpicture}
		\caption{Holographic map $V$ acts with a typical unitary $U$ on $\ell$ and $r$ of the effective theory with some fixed state $|\psi_0\rangle_f$. It post-selects on an auxiliary system $P$ and produces a state in the fundamental description $B$. Similarly, $W$ maps the global charges $Q_\ell$, $Q_r$, and a fixed state $|\wt{\psi}_0\rangle_{f_G}$ to the fundamental charge ${C}$ by a random unitary $U_G$, which post-selects a state of another auxiliary system $P_G$.}\label{fig1}
	\end{center}
\end{figure}
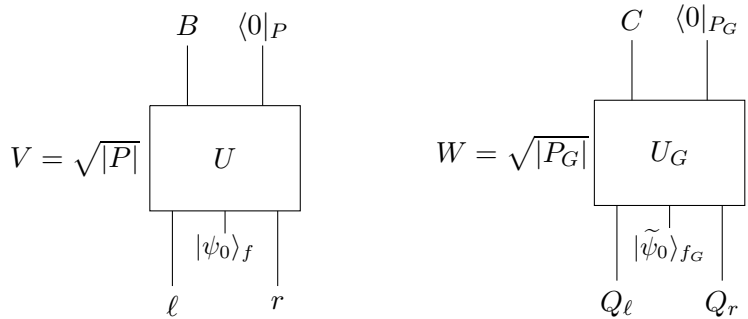

The map $V$ can be defined in terms of a unitary matrix as
\begin{equation}
	V \equiv \sqrt{|P|}\langle0|_PU|\psi_0\rangle_f, 
\end{equation}
where $|\psi_0\rangle\in\calh_f$ and $|0\rangle_P\in\calh_P$ are some fixed states, and $U$ is a typical unitary matrix from the Haar measure. 
The corresponding circuit diagram is depicted in the left panel of figure \ref{fig1}.

Now we introduce global symmetry. In the effective description, each field in $\calh_\ell \otimes \calh_r$ can be charged, so the corresponding charge Hilbert space is $\calh_{Q_\ell} \otimes \calh_{Q_r}$. Charges in the effective description, $Q_\ell$ and $Q_r$, are mapped to the fundamental system $C$ with the Hilbert space $\calh_C$ by another non-isometric map.  
With auxiliary systems $f_G, P_G$, satisfying  
\begin{equation}
	\calh_{Q_\ell}\otimes\calh_{f_G}\otimes\calh_{Q_r}=\calh_{{C}}\otimes\calh_{P_G}, \qquad |Q_\ell||f_G||Q_r|=|{C}||P_G|,
\end{equation}
the non-isometric map for global symmetry is defined as 
\begin{equation}
	W:\calh_{Q_{\ell}}\otimes\calh_{Q_r} \rightarrow \calh_{{C}},
	\quad W \equiv \sqrt{|P_G|}\langle0|_{P_G}U_G|\wt{\psi}_0\rangle_{f_G}.
\end{equation}
Here, $|\wt{\psi}_0\rangle\in\calh_{f_G}$ and $|0\rangle_{P_G}\in\calh_{P_G}$ are some fixed states and $U_G$ is another unitary.
The circuit diagram is depicted in the right panel of figure \ref{fig1}. We denote the non-isometric map
\begin{equation}
	X \equiv V \otimes W,  
\end{equation}
for the combined system. 

Introducing a reference system $L$ with charge $Q_L$ and radiation $R$ with charge $Q_R$, the inner product between two states $|\psi_1\rangle, |\psi_2\rangle\in (\calh_L\otimes\calh_{Q_L}\otimes\calh_\ell\otimes\calh_{Q_\ell})\otimes(\calh_r\otimes\calh_{Q_r}\otimes\calh_R\otimes\calh_{Q_R})$ is
\begin{equation}
	\int dUdU_G \langle\psi_2|(X^\dagger X \otimes I_{LQ_LRQ_R})|\psi_1\rangle = \langle\psi_2|\psi_1\rangle,
\end{equation}
by using the identity of $U(N)$ matrices for the normalized Haar measure,
\begin{equation}
	\int dU\, U_{i_1j_1}U^\dagger_{j'_1i'_1}=\frac{1}{N}\d_{i_1i'_1}\d_{j_1j'_1}
\end{equation}
for $V$ and $W$.  Since we are interested in a typical $U$ and $U_G$, we will average over them. So on average, the non-isometric map preserves the inner product. However, the fluctuation in the inner product is not preserved but bounded.
\begin{equation}\label{inner}
	\int dUdU_G\Big|\langle\psi_2|(X^\dagger X \otimes I_{LQ_LRQ_R})|\psi_2\rangle-\langle\psi_2|\psi_1\rangle\Big|^2 \leq \frac{4}{|B||{C}|}+\frac{4}{|B|}+\frac{4}{|{C}|}. 
\end{equation}
We refer to appendix \ref{appA} for the computation of the bound. The fluctuation is exponentially suppressed in the black hole entropy, $S_{BH}=\log(|B||{C}|)$. We may think of the neutral black hole entropy as $S_N=\log|B|$ and the increase in entropy due to global charges as $S_C=\log|{C}|$. Compared to the case without global charges, which is $2/|B|$ in \cite{Akers:2022qdl}, the upper bound of the fluctuation has increased.

\section{Global symmetry violation}\label{three}

In this section, we compute the entropy of the radiation in the fundamental description and show that it is given by the QES formula in the effective description, which includes the global symmetry-breaking term. The relative entropy between the radiation density matrix and its transform under a global symmetry quantifies the violation of global symmetry. We calculate the relative entropy and find it to be nonzero. We mostly consider $U(1)$ global symmetry for simplicity. 

\subsection{A model of global symmetry}		

In the effective description, the left movers $\ell, Q_\ell$ are to be interpreted as the black hole constituents. They are entangled with the reference systems $L, Q_L$. The right movers $r,Q_r$ are entangled with the radiation system $R,Q_R$. We consider a Hawking state with global symmetry charges,
\begin{equation}\label{psih}
	|\psi_{H}\rangle = |\chi^{in}\rangle_{LQ_L\ell Q_\ell}\otimes |\chi^{out}\rangle_{rQ_rRQ_R},
\end{equation}
where
\begin{align}\label{hawk}
	|\chi^{in}\rangle_{LQ_L\ell Q_\ell} &= \sum_{m,q_L}\sqrt{C_{m,q_L}}\,|m,q_L\rangle_{LQ_L}|m,-q_L\rangle_{\ell Q_\ell}, \nn\\
	|\chi^{out}\rangle_{rQ_rRQ_R} &= \sum_{n,q_R}\sqrt{D_{n,q_R}}\,|n,-q_R\rangle_{rQ_r}|n,q_R\rangle_{RQ_R},
\end{align}
where non-negative constants $C_{m,q_L}$ and $D_{n,q_R}$  sum to one , i.e. $\sum_{n,q_R}D_{n,q_R}=1$, etc. and the ket vectors $|m,q_L\rangle_{LQ_L}$, ... are orthonormal bases. The indices $q_L,q_R$ denote the global charges and $m,n$ label any other eigenstates such as energy in the respective Hilbert spaces. For the $U(1)$ global symmetry, the index runs  $q_L = 0, \pm 1, \pm 2, \dots ,\pm(|Q_L|-1)/2$, if we ignore bound states and $q_R$ runs similarly. One may regard the $U(1)$ as a subgroup of $SU(2)$ so that the states in $\calh_{Q_L}$ are labeled by the $z$ component of the spin $(|Q_L|-1)/2$ representation and similarly for states in other charge spaces. 
We have assumed that the global symmetry is preserved in the effective description, 
\begin{equation}
	q_\ell+q_L=0 , \qquad q_r+q_R=0,
\end{equation} 
where $q_\ell$ and $q_r$ are charges of $Q_\ell$ and $Q_r$, so that Hawking pairs in $(L,\ell)$ and $(r, R)$ have opposite charges. The state \eqref{hawk} implies that the black hole can radiate global charges.  However,  the net charge in the radiation is expected to average to zero
\begin{equation}
	\langle q_R \rangle_{\chi^{out}}=0,
\end{equation}
since the Hawking radiation is thermal \cite{Hawking:1975vcx}.

Let us define maps $V_{p}:\calh_{\ell}\otimes\calh_{r}\rightarrow \calh_{B}$ and $W_p:\calh_{Q_\ell}\otimes\calh_{Q_r}\rightarrow\calh_{{C}}$ such that
\begin{align}\label{doublephase}
	(V_{p}\otimes W_p\otimes I_{LQ_LRQ_R})|\psi_{H}\rangle = \frac{1}{\sqrt{|B||{C}|}}&\sum_{\substack{ m,q_L,\\ n,q_R}}
	\sqrt{C_{m,q_L}D_{n,q_R}}\,|m,q_L\rangle_{LQ_L}|n,q_R\rangle_{RQ_R}\nn\\
	\times&\left(\sum_{b,{c}}\,e^{i\th(m,n,b)+i\a(-q_L,-q_R,{c})}|b,{c}\rangle_{B{C}}\right),
\end{align}
where $e^{i\th(m,n,b)}$ and $e^{i\a(-q_L,-q_R,{c})}$ are random phases due to the chaotic nature of black holes. This model is the extension of the phase model in \cite{Akers:2022qdl}. Here, states $|b,c\rangle_{BC}$ are the orthonormal basis of $\calh_B\otimes \calh_C$. See appendix \ref{singleu} for a discussion of another model.
We use a notation $X_p\equiv V_p\otimes W_p$. 

In this model, we have $\langle\psi_{H}|(X^\dagger_{p}\otimes I_{LQ_LRQ_R})(X_{p}\otimes I_{LQ_LRQ_R})
|\psi_{H}\rangle =1$: the total map, $X_p\otimes I_{LQ_LRQ_R}$, preserves the norm. But the partial map $X_p$ does not
\begin{align}
	\langle m',-q_L';n',-q_R'|&X^\dagger_{p}X_{p}|m,-q_L;n,-q_R\rangle_{\ell Q_\ell rQ_r} \nn\\
	&= \frac{1}{|B||{C}|}\sum_{b,{c}} e^{i(\th(m,n,b)-\th(m',n',b))+i(\a(-q_L,-q_R,{c})- \a(-q_L',-q_R',{c}))} \\
	&= \begin{cases} 1 & \qquad (m,n)=(m',n') ~, ~(q_L,q_R) = (q_L',q_R') \nn\\
		{\cal{O}}\left(\frac{1}{\sqrt{|B|}}\right) & \qquad (m,n)\neq(m',n') ~, ~(q_L,q_R) = (q_L',q_R') \\
		\calo\left(\frac{1}{\sqrt{|{C}|}}\right) & \qquad (m,n)=(m',n')~ , ~(q_L,q_R) \neq (q_L',q_R') \\
		\calo\left(\frac{1}{\sqrt{|B||{C}|}}\right) & \qquad (m,n)\neq(m',n')~ , ~(q_L,q_R) \neq (q_L',q_R'),
	\end{cases}           
\end{align}
where the sums of phases as a random walk are estimated to be of order $\sqrt{|B|}$, $\sqrt{|C|}$ and $\sqrt{|B||{C}|}$, respectively. 
The inner product is approximately preserved for a large $|B|$ or $|C|$. We note that more cases of non-isometry are possible with global symmetry than without it. Nonzero overlaps between different states become significant when $|B|$ or $|C|$ are small.

The density matrix for the Hawking state is mapped by $X_p$ into the fundamental description, 
\begin{equation}\label{density}
	\rho_{LQ_LB{C}RQ_R} = (X_p\otimes I_{LQ_LRQ_R})|\psi_{H}\rangle\langle\psi_{H}|(X_p^\dagger\otimes I_{LQ_LRQ_R}).
\end{equation} 
An outside observer has access to the radiation density matrix $\rho_{RQ_R}$, which is obtained by tracing out the rest 
\begin{align}\label{rhoRp}
	\rho_{RQ_R} &= \tr_{\!LQ_LB{C}}(\rho_{LQ_LB{C}RQ_R}) \nn\\
	&=\sum_{\substack{n,n'\\q_R,q_R'}}\left(\sum_{m,q_L} C_{m,q_L}\sqrt{D_{n,q_R}D_{n',q_R'}}\,\langle m,-q_L, n', -q_R'|X^{\dagger}_{p}X_{p}|m,-q_L, n,-q_R\rangle_{\ell Q_\ell rQ_r} \right)\nn\\
	&\qquad\qquad\times|n,q_R\rangle\langle n',q_R'|_{RQ_R} \nn\\
	&= \frac{1}{|B||{C}|}\sum_{\substack{n,n'\\q_R,q_R'}}\left(\sum_{\substack{m,q_L\\b,{c}}}C_{m,q_L}\sqrt{D_{n,q_R}D_{n',q_R'}}e^{i\th(m,n,b)-i\th(m,n',b)+i\a(-q_L,-q_R,{c})-i\a(-q_L,-q_R',{c})}\right) \nn\\
	&~~~~~~~~~~~~~~~~~~~\times|n,q_R\rangle \langle n',q_R'|_{RQ_R}.
\end{align}
The nonzero components of the off-diagonal charge sector in the density matrix
\begin{equation}\label{rhooff}
	(\rho_{RQ_R})_{(n,q_R)(n',q_R')}, \qquad q_R \neq q_R'
\end{equation}
are global symmetry violating.

\subsection{The quantum extremal surface formula}

Now we carry out the calculations of the QES formula for the radiation.  
The Renyi entropies are given by 
\begin{equation}
	S_n(\rho) = -\frac{1}{n-1}\log\tr(\rho^n).
\end{equation}
The second Renyi of $\rho_{RQ_R}$ is sufficient to see the features of the QES formula,
\begin{align}\label{S2}
	e^{-S_2(\rho_{RQ_R})} = \frac{1}{|B|^2|{C}|^2}&\sum_{\substack{m,m'\\n,n'}}\sum_{\substack{q_L,q_L'\\q_R,q_R'}}C_{m,q_L}C_{m'
		,q_L'}D_{n,q_R}D_{n',q_R'}\nn\\
	\times&\left(\sum_{b,b'}\,e^{i(\th(m,n,b)-\th(m,n',b)+\th(m',n',b')-\th(m',n,b'))}\right)\nn\\
	\times&\left(\sum_{{c},{c}'}\,e^{i(\a(-q_L,-q_R,{c})-\a(-q_L,-q_R',{c})+\a(-q_L',-q_R',{c}')-\a(-q_L',-q_R,{c}'))}\right).
\end{align}
Dominant contributions are when $(n,q_R)=(n',q_R')$, i.e. $\rho_{RQ_R}$ in \eqref{rhoRp} becoming diagonal,  or $(m,q_L;b,{c})=(m',q_L';b',{c}')$. They give
\begin{align}\label{S2final}
	e^{-S_2(\rho_{RQ_R})}& \approx \sum_{n,q_R} D_{n,q_R}^2 + \frac{1}{|B||{C}|}\sum_{m,q_L}C_{m,q_L}^2\nn\\
	&= e^{-S_2(\chi^{out}_{RQ_R})} +  \frac{1}{|B||{C}|}\,e^{-S_2(\chi^{in}_{\ell Q_\ell})},
\end{align}
where $\chi^{out}_{RQ_R}$ and $\chi^{in}_{\ell Q_\ell}$ are reduced density matrices of subsystem $RQ_R$ and $\ell Q_\ell$, respectively in the effective description obtained from \eqref{hawk},
\begin{equation}\label{chioutdef}
	\chi^{out}_{RQ_R} \equiv \tr_{LQ_L\ell Q_\ell rQ_r}\Big(|\psi_H\rangle\langle\psi_H|\Big) 
	= \tr_{rQ_r}\Big(|\chi^{out}\rangle\langle\chi^{out}|_{rQ_rRQ_R}\Big).
\end{equation}
The $\chi^{in}_{\ell Q_\ell}$ is defined similarly. 
We note that the second term of \eqref{S2final} comes from the off-diagonal part of $\rho_{RQ_R}$ and thus breaks the global symmetry. It is essential that the map $X_p$ is non-isometric for symmetry breaking, for if it were not, we would have  $X_p^\dagger X_p=1$ and $\rho_{RQ_R}=(\rho_{\psi_H})_{RQ_R}$, with $(\rho_{\psi_H})_{RQ_R}$ the semiclassical reduced density matrix of $|\psi_H\rangle$ in \eqref{psih}. Thus, the radiation entropy in the fundamental description would be equal to the semiclassical one. There are two more cases of phase cancellation in \eqref{S2}, which will be discussed in a later subsection. 

The result is consistent with the QES formula for generalized entropy,
\begin{equation}\label{Sgen}
	S_2(\rho_{RQ_R}) \approx \text{min}\Big[S_2(\chi^{out}_{RQ_R}), ~\log(|B||{C}|) + S_2(\chi^{in}_{\ell Q_\ell})\Big].
\end{equation}
The first contribution is the effective entropy that describes the early radiation. Late stage is dominated by the second part, which is the area term and the contribution from the interior modes $(\ell,Q_\ell)$: It is reasonable to expect that both $|B|$ and $|C|$ are small at late times. The other interior modes $(r,Q_r)$ are purified by the radiation $(R,Q_R)$: The corresponding contribution vanishes. 
The global symmetry, which is preserved in the early radiation, starts breaking once the second part of \eqref{Sgen} becomes relevant \cite{Hayden:2007cs, Chen:2020ojn}. See \cite{Brown:2024ajk} for a discussion on the evaporation of charged black holes with a gauge field.

One can similarly compute the $n$-th Renyi and obtain the von Neumann entropy. 
For a general map $X$, the density matrix of the radiation is similarly defined as in \eqref{density}. The second Renyi is given by 
\begin{align}
	e^{-S_2(\rho_{RQ_R})} \approx \frac{|P|^2|B|^2|P_G|^2|{C}|^2}{(|P|^2|B|^2-1)(|P_G|^2|{C}|^2-1)}&\left[\left(1-\frac{1}{|P||B|^2}\right)\left(1-\frac{1}{|P_G||{C}|^2}\right)e^{-S_2(\chi^{out}_{RQ_R})}\right.\nn\\
	&+ \left.\left(1-\frac{1}{|P|}\right)\left(1-\frac{1}{|P_G|}\right)\frac{e^{-S_2(\chi^{in}_{\ell Q_\ell})}}{|B||{C}|}\right],
\end{align}
which gives the same result for $|B|,|{C}|,|P|,|P_G|\gg1$. If $|P|=1$ or $|P_G|=1$, the second term in the square bracket becomes zero: the non-isometry of the map $X$ is vital for observing the effects that corresponds to quantum gravity.

\subsection{Relative entropy and global symmetry violation}

We have observed that the second term in \eqref{S2final} breaks the global symmetry. In addition to this, we can also check that the relative entropy does not vanish. 
The relative entropy has been suggested as an order parameter for global symmetry breaking in \cite{Chen:2020ojn}.

\subsubsection{A Renyi relative entropy}

The Renyi relative entropy between $\rho$ and $\sigma$ can be calculated through \cite{Chen:2020ojn,Petz:1986naj}
\begin{equation}
	S(\rho||\sigma) = \lim_{n \rightarrow 1} S_n(\rho||\sigma) \equiv \lim_{n\rightarrow 1}\frac{1}{1-n}\,\tr\Big[\rho\sigma^{n-1} - \rho^n\Big].
\end{equation}

We compare two states: a symmetry transformed state, $\wt{\rho}_{RQ_R}(a) \equiv U_{Q_R}(a)\rho_{RQ_R}U^\dagger_{Q_R}(a)$, and $\rho_{RQ_R}$ in \eqref{rhoRp}, where $U_{Q_R}(a)=e^{ia\hat{Q}_R}$ with $\hat{Q}_R$ the charge operator. The global symmetry transformation $U_{Q_R}$ is a well-defined operator acting on the radiation because of the split property of global symmetries \cite{Harlow:2018tng}. It acts on the basis state as $U_{Q_R}(a)|n,q_R\rangle = e^{iaq_R}|n,q_R\rangle.$ In the effective description, the reduced density matrix of radiation, $\chi^{out}_{RQ_R}$, obtained from \eqref{chioutdef} and \eqref{hawk} is diagonal 
\begin{equation}\label{chiout}
	\chi^{out}_{RQ_R} = \sum_{n,q_R} D_{n,q_R}|n,q_R\rangle\langle n,q_R|,
\end{equation}	
and it is invariant under a global symmetry transformation: $\tilde{\chi}^{out}_{RQ_R}(a) \equiv U_{Q_R}(a)\chi^{out}_{RQ_R}U^\dagger_{Q_R}(a)=\chi^{out}_{RQ_R}$. So, the relative entropy, as well as its Renyi version, is zero in the effective description, 
\begin{equation}
	S(\tilde{\chi}^{out}_{RQ_R}(a)||\,\chi^{out}_{RQ_R})= S_n(\tilde{\chi}^{out}_{RQ_R}(a)||\,\chi^{out}_{RQ_R})=0.
\end{equation}  

In the fundamental description, we expect the non-vanishing relative entropy. Since $\tr(\wt{\rho}_{RQ_R}(a)^n)=\tr(\rho_{RQ_R}^n)$, the second Renyi relative entropy is given by
\begin{align}\label{2renyrel}
	S_2(\wt{\rho}_{RQ_R}(a)||\,\rho_{RQ_R}) &= \sum_{q_R,q_R'}\Big(1-e^{ia(q_R-q_R')}\Big) \Big[\tr_R(\rho_{RQ_R})_{q_Rq_R'}(\rho_{RQ_R})_{q_R'q_R}\Big] \nn\\
	&\approx \sum_{q_R,q_R'}\Big(1-e^{ia(q_R-q_R')}\Big)\left[\Big(\sum_{n,n'}D_{nq_R}D_{n'q_R'}\Big)\frac{1}{|B||{C}|}\sum_{m,qL}C_{m,q_L}^2\right]\nn\\
	&=\left[1-\sum_{q_R,q_R'}e^{ia(q_R-q_R')}\sum_{n,n'}D_{nq_R}D_{n'q_R'}\right]\frac{1}{|B||{C}|}e^{-S_2\big(\chi^{in}_{\ell Q_\ell}\big)}
\end{align}
where \eqref{S2} has been used. We note that the diagonal part of $\rho_{RQ_R}$ yields zero and only the off-diagonal part of $\rho_{RQ_R}$ contributes to the relative entropy. That is in the first case of phase cancellation, $\{q_R=q_R'\}$, the above Renyi relative entropy $S_2(\tilde{\rho}_{RQ_R}||\rho_{RQ_R})=0$. The factor in the first line, $\tr_R(\rho_{RQ_R})_{q_Rq_R'}(\rho_{RQ_R})_{q_R'q_R}$, corresponds to the squared amplitude for charge transition. For a small $a$, up to quadratic order 
\begin{align}
	S_2(\wt{\rho}_{RQ_R}(a)||\,\rho_{RQ_R}) &\approx \left[ \frac{a^2}{2}\sum_{n,n'}\sum_{q_R,q_R'}(q_R^2-2q_Rq_R'+q_R'^2)D_{nq_R}D_{n'q_R'}\right]\frac{1}{|B||{C}|}e^{-S_2\big(\chi^{in}_{\ell Q_\ell}\big)}\nn\\
	&= a^2\left(\langle q_R^2\rangle_{\chi^{out}_{RQ_R}} - \langle q_R \rangle^2_{\chi^{out}_{RQ_R}}\right)\frac{1}{|B||{C}|}e^{-S_2\big(\chi^{in}_{\ell Q_\ell}\big)}.
\end{align}
Since $\chi^{out}_{RQ_R}$ in \eqref{chiout} is the effective density matrix, the expectation value $\langle q_R\rangle_{\chi^{out}_{RQ_R}}=0$ \cite{Hawking:1976ra, Hawking:1982dj}. However, $\langle q_R^2\rangle_{\chi^{out}_{RQ_R}}$ does not vanish. 
We have found that the second Renyi relative entropy of a global symmetry transformation has a nontrivial contribution. Therefore, the global symmetry is violated.

The $n$-th Renyi relative entropy is calculated similarly.
\begin{equation}
	S_n(\tilde{\rho}||\rho) = \frac{a^2\langle q_R^2 \rangle_{\chi^{out}_{RQ_R}}}{1-n}\frac{1}{\left(|B||C|\right)^{n-1}}e^{-(n-1)S_n\big(\chi^{in}_{\ell Q_\ell} \big)}
\end{equation}
It goes to infinity in the limit of $n\rightarrow 1^-$.

\subsubsection{The sandwiched Renyi relative entropy}

Although the Renyi relative entropy formula in the previous section can be used to show global symmetry breaking, it is more appropriate to consider an operationally meaningful formula, namely the sandwiched Renyi relative entropy \cite{Wilde:2013bdg, Muller-Lennert:2013liu}, defined by 
\begin{equation}\label{sand1}
	\tilde{S}_n(\rho||\sigma) \equiv \frac{1}{n-1}\log\tr \left[\left(\sigma^{\frac{1-n}{2n}}\rho\,\sigma^{\frac{1-n}{2n}}\right)^n \right].
\end{equation}
This formula satisfies the data-processing inequality for a broader range of $n\in[\frac{1}{2},1) \cup (1,\infty)$ than the traditional Petz quasi-entropy, $D_n(\rho||\sigma)\equiv \frac{1}{n-1}\log\tr\left(\rho^n\sigma^{1-n}\right)$. Also, $\tilde{S}_n(\rho||\sigma) \leq D_n(\rho||\sigma)$ for all $n>1$. \cite{Frank:2013rov, Muller-Lennert:2013liu,Wilde:2013bdg} 

We first calculate $\tr[(\rho_{RQ_R})^p]$. With a notation $\rho_{RQ} \equiv \rho$ and $q_R\equiv q$, the radiation density matrix \eqref{rhoRp} is  
\begin{equation}
	\rho_{(nq)_1,(nq)_2} = \frac{1}{ |B||C|}\left(\sum_{\substack{m,q_L\\b,{c}}}C_{m,q_L}\sqrt{D_{n_1,q_{1}}D_{n_2,q_{2}}}e^{i\th(m,n_1,b)-i\th(m,n_2,b)+i\a(-q_L,-q_{1},{c})-i\a(-q_L,-q_{2},{c})}\right).
\end{equation}
Then 
\begin{equation}\label{rhop}
	\tr(\rho^p) = \frac{1}{\left(|B||C|\right)^p}\sum_{\vec{n},\vec{q}}\sum_{\substack{\vec{m},\vec{q}_L\\\vec{b},\vec{c}}}\left(\prod_{j=1}^pC_{m_j,q_{Lj}}D_{n_jq_j}e^{i\Theta(m_j,n_j,b_j)}e^{i\cala(q_{Lj},q_j,c_j)}\right),
\end{equation}
where the vectors have $p$ components, e.g. $\vec{n}=(n_1, n_2, ..., n_p)$, and we have defined $\Theta(m_i,n_i,b_i)\equiv \th(m_i,n_i,b_i)-\th(m_i,n_{i+1},b_i)$ and $\cala(q_{Li}, q_i, c_i)\equiv \a(-q_{Li},-q_i, c_i)-\a(-q_{Li},-q_{i+1},c_i)$ with $n_{p+1}\equiv n_1$ and $q_{p+1}\equiv q_1$. Dominant contributions come from the two cases of phase cancellation: $\{n_i=n, q_i=q\}$ for all $i$ or $\{m_i=m, q_{Li}=q_L, b_i=b, c_i=c\}$ for all $i$. So \eqref{rhop} is given by
\begin{align}\label{rhopf}
	\tr (\rho^p) &\approx \sum_{n,q} (D_{n,q})^p+ \frac{1}{\left(|B||C|\right)^{p-1}}\sum_{m,q_L} (C_{m,q_L})^p \nn\\
	&= e^{-(p-1)S_p\big(\chi^{out}_{RQ_R}\big)} + \frac{1}{\left(|B||C|\right)^{p-1}}e^{-(p-1)S_p\big(\chi^{in}_{\ell Q_\ell}\big)}.
\end{align}

Now we compute $\tr[\left(\rho^m\tilde{\rho}\rho^m\right)^n]$ with $\tilde{\rho}\equiv U(a)\rho\, U^\dagger(a)$. Writing $\rho_{(nq)_i,(nq)_j}$ as $\rho_{t_it_j}$ and $q_{i,j} \equiv q_{i}-q_{j}$,   it can be written
\begin{align}
	\tr[\left(\rho^m\tilde{\rho}\rho^m\right)^n] &= \sum_{\vec{t}}\prod_{s=1}^n\left[ (\rho^m)_{t_{3s-2},t_{3s-1}}e^{ia(q_{3s-1} -q_{3s})}\rho_{t_{3s-1},t_{3s}}(\rho^m)_{t_{3s},t_{3s+1}}\right] \nn\\
	&= \sum_{\vec{t}}\prod_{s=1}^n \left[e^{ia\left( q_{3s-1,3s}\right)}(\rho^m)_{t_{3s-2},t_{3s-1}}\rho_{t_{3s-1},t_{3s}}(\rho^m)_{t_{3s},t_{3s+1}}\right],
\end{align}
where $t_{3n+1} \equiv t_1$. For the parameter $a$ small,  we obtain up to quadratic in $a$
\begin{align}
	&\tr[\left(\rho^m\tilde{\rho}\rho^m\right)^n] \nn\\
	&= \sum_{\vec{t}}\left[1+ia\left(\sum_{s=1}^n q_{3s-1,3s}\right)-\frac{a^2}{2}\left(\sum_{s=1}^nq_{3s-1,3s}\right)^2\right]\prod_{s=1}^n (\rho^m)_{t_{3s-2},t_{3s-1}}\rho_{t_{3s-1},t_{3s}}(\rho^m)_{t_{3s},t_{3s+1}}.
\end{align}
This approximation is valid for $|n|<1$. Otherwise, higher order terms in $a$ become important since they involve higher powers of $n$. So we analytically continue in $n$.  
The phases of the product of $\rho$'s cancel for the same cases as in $\tr(\rho^p)$ computed in \eqref{rhopf}. In the first case, all $q_i's$ are equal so the contribution is 
\begin{equation}\label{cont1}
	\tr[\left(\rho^m\tilde{\rho}\rho^m\right)^n] = \sum_{n,q}(D_{n,q})^{(2m+1)n}.
\end{equation}
In the second case, we note that $\sum_{n,q} qD_{nq}=\sum_{n,q} q\,(\chi^{out}_{RQ_R})_{nq, nq}=\langle q\rangle_{\chi^{out}_{RQ_R}}=0$ as in the previous section. So the linear terms in $a$ vanish. For the same reason, among terms quadratic in $a$, only $q_i^2$ terms survive.  They contribute
\begin{equation}\label{cont2}
	\tr[\left(\rho^m\tilde{\rho}\rho^m\right)^n] = \left(1-na^2\langle q^2\rangle_{\chi^{out}_{RQ_R}}\right)\frac{1}{\left(|B||C|\right)^{(2m+1)n-1}}\sum_{j,q_L}(C_{j,q_L})^{(2m+1)n}.
\end{equation}
Combining \eqref{cont1} and \eqref{cont2} with $m=(1-n)/2n$, we find
\begin{equation}\label{sandtr}
	\tr\left[\left(\rho^{\frac{1-n}{2n}}\tilde{\rho}\rho^{\frac{1-n}{2n}}\right)^n\right] =\begin{cases} &1 \nn\\
		& 1-na^2\langle q^2\rangle_{\chi^{out}_{RQ_R}},  \qquad |n|<1.
	\end{cases}
\end{equation}
The expression in the second line corresponds to quantum gravity effects. So the sandwiched Renyi relative entropy is either zero or 
\begin{align}\label{sand}
	\tilde{S}_n(\tilde{\rho}||\rho) &= \frac{1}{n-1}\log\left(1-na^2\langle q^2\rangle_{\chi^{out}_{RQ_R}}\right) \nn\\
	&\approx \frac{n}{1-n}a^2\langle q^2\rangle_{\chi^{out}_{RQ_R}}.
\end{align}
Thus, when there are no non-perturbative gravity effects, the relative entropy is  
\begin{equation}
	\tilde{S}(\tilde{\rho}||\rho) = 0,
\end{equation}
effectively. Once non-perturbative gravity effects are taken into account, the relative entropy diverges 
\begin{equation}
	\tilde{S}(\tilde{\rho}||\rho) = \lim_{n\rightarrow 1^-}\tilde{S}_n(\tilde{\rho}||\rho) \rightarrow +\infty
\end{equation}
The two states $\tilde{\rho}$ and $\rho$ are completely different. This shows a global symmetry violation in quantum gravity. 

Another quantity to distinguish two states is the fidelity, 
\begin{equation}
	F(\rho, \sigma) \equiv \Big(\tr|\sqrt{\rho}\sqrt{\sigma}| \Big)^2,
\end{equation} 
where $|A| = \sqrt{A^\dagger A}$. It attains its maximum value 1 if and only if the two states are identical and vanishes for mutually orthogonal states. It is related to the sandwiched Renyi relative entropy \eqref{sand1}, 
\begin{equation}
	\tilde{S}_{1/2}(\rho||\sigma) = -2\log\tr\left[\left(\sqrt{\sigma}\rho\sqrt{\sigma}\right)^{\frac{1}{2}}\right]= -\log F(\rho,\sigma),
\end{equation}
\cite{Muller-Lennert:2013liu}.
In our case, the fidelity between $\tilde{\rho}$ and $\rho$ is found from \eqref{sand}, 
\begin{equation}
	F(\tilde{\rho},\rho) = \exp\left(-a^2\langle q^2\rangle_{\chi^{out}_{RQ_R}} \right).
\end{equation}
The two states become more distinguishable, as the global symmetry parameter $a$ or the expectation value of charge squared increases.

\subsection{Comments on black hole remnant}\label{sec:more}

We discuss two more cases of phase cancellation in \eqref{S2}. The first case is when the global charges are preserved in the radiation: $q_R=q_R'$ with $(m,b)=(m',b')$. The second corresponds to the conditions: $n=n'$ with $(q_L,{c})=(q_L',{c}')$ and the global charges in the radiation are not necessarily conserved. Their contributions to the second Renyi are
\begin{align}\label{S2inter}
	e^{-S_2(\rho_{RQ_R})} &\approx  \frac{1}{|B|}\sum_{\substack{m,n\\n'}}\sum_{\substack{q_L,q_R\\q_L'}}C_{m,q_L}C_{m,q_L'}D_{n,q_R}D_{n',q_R}+ \frac{1}{|{C}|}\sum_{\substack{m,n\\m'}}\sum_{\substack{q_L,q_R\\q_R'}} C_{m,q_L}C_{m',q_L}D_{n,q_R}D_{n,q_R'}\nn\\
	&= \frac{1}{|B|}\,e^{-S_2\big(\chi^{in}_{\ell }\otimes\,\chi^{out}_{Q_R}\big)} + \frac{1}{|{C}|}\, e^{-S_2\big(\chi^{in}_{Q_\ell}\otimes\,\chi^{out}_{R}\big)}\nn\\
	&\approx \frac{1}{|B|}\left(\frac{1}{|\ell||Q_R|}\right) + \frac{1}{|{C}|}\left(\frac{1}{|Q_\ell||R|}\right),
\end{align}   
where $|\ell|,|R|,|Q_\ell|,|Q_R|$ are dimensions of the respective spaces and
in the last line, we have assumed $C_{m,q_L}\sim 1/|\ell||Q_\ell|$ and $D_{n,q_R}\sim 1/|R||Q_R|$ for simplicity.

These contributions can be understood as intermediate stages of black hole evaporation. The first term in \eqref{S2inter} corresponds to the form of the generalized entropy, in which the island is formed for the neutral modes while the charge modes are still in the early stage. This contribution could come from black hole remnants \cite{Susskind:1995da}, which would have small $|B|$ with large charge degeneracy, $|Q_r|=|Q_R|$ in the effective description. They preserve the global symmetry, but violate the Bekenstein bound \cite{Banks:2010zn}. We discuss below how their contribution is suppressed at very late times. 
In the second term of \eqref{S2inter}, the charge modes are in the late phase with their island, while the neutral modes remain in the early phase.

For $|B|\ll|C|$, the remnant contribution is dominant so comparing the first term of \eqref{S2inter} with the last term of \eqref{S2final},  we have 
\begin{align}\label{S2comp1}
	S_2(\rho_{RQ_R}) & \approx \log |B| + \text{min}\Big[S_2\big(\chi^{in}_{\ell}\otimes\chi^{out}_{Q_R}\big), ~\log |{C}|+S_2\big(\chi^{in}_{\ell Q_\ell}\big)\Big] \nn\\
	& = \log |B| + S_2(\chi^{in}_\ell) + \text{min}\Big[S_2\big(\chi^{out}_{Q_R}\big), ~\log |{C}| +S_2\big(\ell Q_\ell|\,\ell\big)\Big]  ,
\end{align} 
where $S_2(\ell Q_\ell|\,\ell) \equiv S_2(\chi^{in}_{\ell Q_\ell}) - S_2(\chi^{in}_\ell)$ is the conditional Renyi entropy.\footnote{Although this definition of the conditional Renyi entropy does not satisfy the monotonicity of the conditional entropy \cite{teixeira2012conditional}, we won't worry about it. } The above expression can be regarded as a QES formula for global charge degrees of freedom. We expect a similar Page curve \cite{Page:1993wv} for global charges as seen in neutral black holes. 
For late times, we can approximate $S_2(\chi^{out}_{Q_R})\approx \log|Q_R|$, $S_2(\chi^{in}_{\ell})\approx\log|\ell|$, and $S_2(\chi^{in}_{\ell Q_\ell})\approx \log (|\ell||Q_\ell|)$. Since the non-isometric map eventually makes $|C|$ small, we will have $\log|Q_R|>\log(|C||Q_\ell|)$. Therefore, at very late times, the entropy comes from the last terms in \eqref{S2comp1}.

For $|C|\ll|B|$, we can compare the second term in \eqref{S2inter} and \eqref{S2final},   
\begin{align}
	S_2(\rho_{RQ_R})  &\approx \log|C| +S_2(\chi^{in}_{Q_\ell})+ \text{min}\Big[S_2\big(\chi^{out}_{R}\big), ~\log |B| + S_2(\ell Q_\ell|\,Q_\ell)\Big] .
\end{align} 
This can be interpreted as a QES formula for neutral degrees of freedom.  

At sufficiently late times, the black hole degeneracy will be smaller than the radiation degeneracy: both $|B|\ll|R|$ and $|{C}|\ll|Q_R|$ are satisfied. Therefore, the conditions
\begin{equation}
	|Q_R|>|{C}||Q_\ell| , \qquad |R|>|B||\ell|
\end{equation}
hold and the contributions in (\ref{S2inter}) become negligible.

\section{Discussion}

We have modeled black holes as non-isometric codes for global symmetries to demonstrate that global symmetries are violated in quantum gravity. The radiation density matrix in the fundamental description has global symmetry violating components. 

We have calculated the Renyi entropies and relative entropies for the radiation system with global charges. The Renyi entropies include the off-diagonal contributions of the density matrix and are consistent with the QES formula. Additionally, the Renyi versions of relative entropy for a global symmetry transformed state are calculated and found to be nonzero, indicating a violation of global symmetry. Specifically, the sandwiched Renyi relative entropy is utilized and the fidelity was computed. The non-isometry of the map $V$ and $W$ plays a vital role in the derivation of the results, yielding a nonzero relative entropy. It in fact blows up, when non-perturbative gravity effects are present. We have also discussed the entropies of the remnant-like configurations and the suppression of their contribution through the QES formula.

Related to this work, further questions could be raised regarding more general global symmetries such as discrete and $p$-form global symmetries, the completeness conjecture \cite{Polchinski:2003bq,Arkani-Hamed:2006emk}, the reconstruction of global charge states, and relations to complexity \cite{Akers:2022qdl}.

\subsection*{acknowledgments}
	
	This work is supported by the Grant No. RS-2023-00277184 of the National Research Foundation of Korea.

\appendix
\section{Calculation of the fluctuation in the inner product}\label{appA}

We give details for the calculation of the fluctuation in the inner product \eqref{inner}.

Applying the integral formula  
\begin{align}
	\int dU\, U_{i_1j_1}U_{i_2j_2}U^\dagger_{j'_1i'_1}U^\dagger_{j'_2i'_2}  = \frac{1}{N^2-1}&\left[ \d_{i_1i'_1}\d_{i_2i'_2}\d_{j_1j'_1}\d_{j_2j'_2} + \d_{i_1i'_2}\d_{i_2i'_1}\d_{j_1j'_2}\d_{j_2j'_1}\right.\nn\\
	&- \left.\frac{1}{N}\Big(\d_{i_1i'_1}\d_{i_2i'_2}\d_{j_1j'_2}\d_{j_2j'_1} + \d_{i_1i'_2}\d_{i_2i'_1}\d_{j_1j'_1}\d_{j_2j'_2}\Big)\right]
\end{align}
twice for averaging over $U$ and $U_G$, the first term of the fluctuation in the inner product \eqref{inner} is computed as 
\begin{align}
	\int dUdU_G \Big|\langle\psi_2|(V^\dagger V\otimes W^\dagger W\otimes &I_{LQ_LRQ_R})|\psi_1\rangle\Big|^2 \nn\\
	= \frac{|P|^2|P_G|^2}{(|P|^2|B|^2-1)(|P_G|^2|{C}|^2-1)}&\left[\left(|B|^2-\frac{|B|}{|P||B|}\right)\left(|{C}|^2-\frac{|{C}|}{|P_G||{C}|}\right)\big|\langle\psi_2|\psi_1\rangle\big|^2\right. \nn\\
	&+\left(|B|^2-\frac{|B|}{|P||B|}\right)\left(|{C}|-\frac{|{C}|^2}{|P_G||{C}|}\right)\tr(\psi_{1S}\psi_{2S}) \nn\\
	&+ \left(|B|-\frac{|B|^2}{|P||B|}\right)\left(|{C}|^2-\frac{|{C}|}{|P_G||{C}|}\right)\tr(\psi_{1Q}\psi_{2Q})\nn\\ 
	&+\left. \left(|B|-\frac{|B|^2}{|P||B|}\right)\left(|{C}|-\frac{|{C}|^2}{|P_G||{C}|}\right)\tr(\psi_{1}\psi_{2})\right],
\end{align}
where $\psi_S\equiv \tr_{Q_{\ell}Q_r}\,\psi$ and $\psi_Q \equiv \tr_{\ell r}\,\psi$ are reduced density matrices.  
Then, the fluctuation is given by
\begin{align}\label{fluc2}
	\int &dUdU_G \Big|\langle\psi_2|(V^\dagger V\otimes W^\dagger W\otimes I_{LQ_LRQ_R})|\psi_1\rangle -\langle\psi_2|\psi_1\rangle\Big|^2\nn\\
	=& \frac{|P||P_G|}{(|P|^2|B|^2-1)(|P_G|^2|{C}|^2-1)}\Big[|B||{C}|(|P|-1)(|P_G|-1)\,\tr(\psi_1\psi_2) \nn\\
	&+ |B|(|P|-1)(|P_G||{C}|^2-1)\,\tr(\psi_{1Q}\psi_{2Q})+ |{C}|(|P||B|^2-1)(|P_G|-1)\,\tr(\psi_{1S}\psi_{2S})\Big]\nn\\
	&- \frac{|P|^2|B|^2(|P_G|-1)+ |P_G|^2|{C}|^2(|P|-1)-|P||P_G|+1}{(|P|^2|B|^2-1)(|P_G|^2|{C}|^2-1)}\big|\langle\psi_1|\psi_2\rangle\big|^2
\end{align}
When $|P_G|=1$, the right-hand side becomes identical to the expression without global charges in \cite{Akers:2022qdl}. When both $|P|=|P_G|=1$, the above expression vanishes as it should be since it corresponds to the absence of null states.

Using the fact that $\frac{1}{x-1}\geq\frac{2}{x}$ for $x\geq 2$ and $\tr(\psi_1\psi_2)\leq1$, the upper bound for the fluctuation \eqref{fluc2} is
\begin{equation}
	\int dUdU_G \Big|\langle\psi_2|(V^\dagger V\otimes W^\dagger W\otimes I_{LQ_LRQ_R})|\psi_1\rangle -\langle\psi_2|\psi_1\rangle\Big|^2 \leq \frac{4}{|B||{C}|}+\frac{4}{|B|}+\frac{4}{|{C}|}
\end{equation}
for $|B|\gg1$ and $|{C}|\gg1$. 

\section{A single unitary model}\label{singleu}
The model \eqref{doublephase} involves two independent phases $\th$ and $\a$. The states of a subsystem are labeled by two indices, e.g. the right-moving subsystem $rQ_r$ by $(n,q_r)$, which denote energy and global charge. They are scrambled independently via two unitaries $V_p$ and $W_p$ in a black hole. This is motivated by the fact that a global symmetry charge does not alter the energy of a given state, such as an open string state with the Chan-Paton factor. Since there is no gauge field associated with global symmetries, global symmetry charges are non-dynamical and do not interact dynamically. 

Alternatively, one may consider a single unitary map, which acts on all four Hilbert spaces in the effective description, 
$$\tilde{V}: \calh_\ell\otimes\calh_r\otimes\calh_{Q_\ell}\otimes\calh_{Q_r} \rightarrow \calh_B\otimes\calh_C$$ with extra systems $\tilde{f}$ and $\tilde{P}$ such that 
$$\calh_\ell\otimes\calh_r\otimes\calh_{Q_\ell}\otimes\calh_{Q_r}\otimes\calh_{\tilde{f}}=\calh_B\otimes\calh_C\otimes\calh_{\tilde{P}}, \qquad |\ell||r||Q_\ell||Q_r||\tilde{f}|=|B||C||\tilde{P}|.$$ 
This mapping combines energy and global symmetry charge within a black hole through a single unitary transformation. 
It can be written in terms of a unitary $\tilde{U}$, 
$$\tilde{V} = \sqrt{\tilde{P}}\langle 0|_{\tilde{P}}\tilde{U}|\psi_0\rangle_{\tilde{f}},$$
where $|0\rangle_{\tilde{P}}\in\calh_{\tilde{P}}$ and $|\psi_0\rangle_{\tilde{f}}\in\calh_{\tilde{f}}$ are some fixed states. 

Specifically, a single phase model $\tilde{V}_p$ is given by its action on the Hawking state,
\begin{align}
	(\tilde{V}_{p}\otimes I_{LQ_LRQ_R})|\psi_{H}\rangle = \frac{1}{\sqrt{|B||C|}}&\sum_{\substack{ m,q_L,\\ n,q_R}}
	\sqrt{C_{m,q_L}D_{n,q_R}}\,|m,q_L\rangle_{LQ_L}|n,q_R\rangle_{RQ_R}\nn\\
	\times&\left(\sum_{b,{c}}\,e^{i\th(m,n,b;-q_L,-q_R,c)}|b,c \rangle_{BC}\right),
\end{align}
where the phase $\th$ depends on both energy and global charge. The inner product in this model is
\begin{align}
	\langle m',-q_L';n',-q_R'|&\tilde{V}^\dagger_{p}\tilde{V}_{p}|m,-q_L;n,-q_R\rangle_{\ell Q_\ell rQ_r} \nn\\
	&= \frac{1}{|B||C|}\sum_{b,c} e^{i(\th(m,n,b;-q_L,-q_R,c)-\th(m',n',b;-q_L',-q_R',c))} \\
	&= \begin{cases} 1 & \qquad (m,n)=(m',n') ~, ~(q_L,q_R) = (q_L',q_R') \nn\\
		\calo\left(\frac{1}{\sqrt{|B||{C}|}}\right) & \qquad \text{otherwise}
	\end{cases}           
\end{align} 
The radiation density matrix becomes
\begin{align}\label{rhoRp}
	\rho_{RQ_R} &= \tr_{\!LQ_LB{C}}(\rho_{LQ_LB{C}RQ_R}) \nn\\
	&= \frac{1}{|B||{C}|}\sum_{\substack{n,n'\\q_R,q_R'}}\left(\sum_{\substack{m,q_L\\b,c}}C_{m,q_L}\sqrt{D_{n,q_R}D_{n',q_R'}}e^{i\th(m,n,b; -q_L,-q_R,c)-i\th(m,n',b; -q_L,-q_R',c)}\right) \nn\\
	&~~~~~~~~~~~~~~~~~~~\times|n,q_R\rangle \langle n',q_R'|_{RQ_R}.
\end{align}
The second Renyi entropy is obtained from 
\begin{align}
	\tr\left(\rho_{RQ_R}^2\right) = &\frac{1}{|B|^2|{C}|^2}\sum_{\substack{m,m'\\n,n'}}\sum_{\substack{q_L,q_L'\\q_R,q_R'}}C_{m,q_L}C_{m'
		,q_L'}D_{n,q_R}D_{n',q_R'}\nn\\
	\times&\left(\sum_{b,b'}\sum_{{c},{c}'}\,e^{i(\th(m,n,b;-q_L,-q_R,{c})-\th(m,n',b;-q_L,-q_R',{c})+\th(m',n',b';-q_L',-q_R',{c}')-\th(m',n,b';-q_L',-q_R,{c}'))}\right).
\end{align}
The QES formula and relative entropy results of this model are the same as before but the non-isometry properties are different. Only two cases of phase cancellation for $\tr\rho_{RQ_R}^2$ exist, where $(n,q_R)=(n',q_R')$ or $(m,q_L;b,c)=(m',q_L';,b',c')$. Thus, the discussion of section \ref{sec:more} is irrelevant in this model. 

\bibliographystyle{jhep}
\bibliography{ref}

\end{document}